\documentclass[aps,prl,showpacs,twocolumn,amsmath,10pt,superscriptaddress,floatfix,nofootinbib,a4paper]{revtex4-1}
\begin{document}
\title{Building up the spectrum of pentaquark states as hadronic molecules}

\author{Bing-Song Zou}
\email{zoubs@itp.ac.cn}

\affiliation{CAS Key Laboratory of Theoretical Physics, Institute of Theoretical Physics, Chinese Academy of Sciences,\\  Zhong Guan Cun East Street 55, Beijing 100190, China}
\affiliation{School of Physical Sciences, University of Chinese Academy of Sciences, Beijing 100049, China}
\affiliation{School of Physics, Central South University, Changsha 410083, China}

\begin{abstract}
I give a brief comment on the LHCb's discovery of $P_c$ and $P_{cs}$ pentaquark states. These pentaquark states not only reveal a new kind of hadronic molecules, but also shed new light on the possible structure of pentaquark components in baryons, hence may play very important role for understanding un-quenching dynamics and the whole baryon spectroscopy.  
\end{abstract}

\maketitle

Following the observation of three narrow $N^*$ pentaquark states with hidden charm, $P_c(4312)$, $P_c(4440)$ and $P_c(4457)$~\cite{Aaij:2015tga, Aaij:2019vzc}, very recently, the LHCb collaboration reported the first evidence for their strange $\Lambda^*$ counterparts, $P_{cs}(4459)$,
in the $J/\psi\Lambda$ invariant mass distribution~\cite{Aaij:2020gdg}. While the $P_c(4312)^+$ sits 6~MeV below the $\bar{D}^0\Sigma_c^+$ threshold, $P_c(4440)$ and $P_c(4457)$ sit 20~MeV and 3~MeV below the $\bar{D}^{*0}\Sigma_c^+$ threshold, respectively, the newly observed $P_{cs}(4459)$ is just about 19~MeV below the $\bar{D}^{*0}\Xi_c^0$ threshold.  These narrow $N^*$ and $\Lambda^*$ pentaquark states are consistence with the corresponding theoretical prediction~\cite{Wu:2010jy} based on the hadronic molecular picture within theoretical uncertainties~\cite{Wang:2011rga,Yang:2011wz,Wu:2012md,Xiao:2021rgp}.

There has been a long history of searching for the pentaquark states~\cite{Guo:2017jvc,Chen:2016qju}.
The first pentaquark state was in fact predicted and observed several years before the quark model was proposed. 
As a possible meson-baryon molecule composed of one kaon and one nucleon, the $\Lambda(1405)$ resonance was predicted by Dalitz and Tuan in 1959~\cite{ Dalitz:1959dn} and observed in the hydrogen bubble chamber at Berkeley in 1961. 
However, with the quark model developed in the early 1960s, the $\Lambda(1405)$ resonance could also be ascribed as an excited state of a three-quark (uds) system with one quark in an orbital P-wave excitation. After about 50 years' fighting on whether it is a ($uds$)-system or ($udsq\bar q$ system of the $\bar KN$ type, until 2010, the Particle Data Group still claimed that ``The clean $\Lambda_c$ spectrum has in fact been
taken to settle the decades-long discussion about the nature of the
$\Lambda(1405)$ -- true 3-quark state or mere $\bar KN$ threshold
effect? -- unambiguously in favor of the first interpretation." 
A similar situation happened for the $N^*(1535)$. It was proposed to be a $K\Sigma$-$K\Lambda$ dynamical generated state~\cite{Kaiser:1995eg}  
with further strong evidence of $qqqs\bar s$ configuration~\cite{Liu:2005pm}. But it is hard to convince people who believe in 3-quark picture.
The difficulty to pin down the nature of these
baryon resonances is that the predicted states from various
models are around the same energy region and there are
always some adjustable ingredients in each model to fit
the experimental data. 
 
A possible solution to settle down this issue was proposed in Ref.\cite{Wu:2010jy} to find the brothers or sisters
of the $\Lambda(1405)$ and $N^*(1535)$ in the hidden charm sector, {\sl i.e.}, replacing $q\bar q$ in $\Lambda(1405)$ and $s\bar s$ in $N^*(1535)$ by $c\bar c$. The study of the interactions between various charmed mesons and
charmed baryons within the framework of the coupled channel unitary approach with the local hidden gauge
formalism led to the prediction of $D^{(*)}\Sigma_c$ and $D^{(*)}\Xi_c$ molecular states, which may be found through their decays to $J/\psi p$ and $J/\psi\Lambda$, respectively. Indeed the LHCb Collaboration observed these $P_c$ and $P_{cs}$ states just below these thresholds with these decay modes. These states definitely cannot be accommodated
by the conventional 3q quark models. Their discovery starts a new era for exploring pentaquark spectrum and has triggered a great renaissance on the study of pentaquark structures.  While these pentaquark states give strong support to the hadronic molecular picture due to their perfect match with its predictions, other possible configurations, such as compact pentaquark states or kinetic effects were also discussed as reviewed in Refs.\cite{Guo:2017jvc,Chen:2016qju}.

A survey of heavy-antiheavy hadronic molecules is given in Ref.~\cite{Dong:2021juy}. All possible combinations of hadron pairs of the S-wave singly-charmed mesons and baryons as well as the narrow P-wave charmed mesons are considered. In total, 229 molecular states are predicted, including 18 $P_c$  and 25 $P_{cs}$ pentaquark states. They are formed by attractive interaction mainly due to light vector meson exchange between charmed hadrons. The favourable decay modes to search for the $P_c$ and $P_{cs}$ states are the $\bar{D}^{(*)}\Lambda_c$ and $\bar{D_s}^{(*)}\Lambda_c$, respectively, in addition to the $J/\psi p$ and $J/\psi\Lambda$~\cite{Guo:2017jvc,Xiao:2021rgp}. The resonant peaks for the $P_c$ and $P_{cs}$ states should appear very close to the corresponding  $D^{(*)}\Sigma_c^{(*)}$ and $D^{(*)}\Xi_c^{(')}$ thresholds, respectively, either below the thresholds as bound states or just at the thresholds as virtual states depending on the details of the strength of the relevant attractive interactions~\cite{Dong:2020hxe}. Similar spectra are expected for their strange and beauty partners~\cite{Guo:2017jvc,Dong:2020hxe}. The LHCb observation of the $P_c$ and $P_{cs}$ states is just a start for the great journey to building up the spectrum of  pentaquark states. 

Building up and understanding the spectrum of pentaquarks is crucial for understanding the whole baryon spectroscopy. Even for the lightest baryon, the proton, to explain its $\bar{u}/\bar{d}$ asymmetry, its strangeness and its spin crisis, a mixture of about 30\% pentaquark component is needed~\cite{Brodsky:1996hc,Zou:2005xy}. To achieve a better understanding of baryon spectroscopy and baryon structure, it is necessary to
extend the classical quenched 3-quark models to include pentaquark
components, {\sl i.e.}, to allow for the mixing of $q^3 \leftrightarrow q^4\bar q$.
The low-lying  $\Omega^*$ states with negative parity could play a
unique role for checking quenched $qqq$ quark models and unquenched
$q^3 \leftrightarrow q^4\bar q$ quark models. Because all three
valence quarks in a $\Omega^*$ state are strange ones, then the
five-quark configurations with a light $q\bar q$ pair must play a
very special role in its properties. Since the $q\bar q$ pair has a
different flavor with valence strange quarks, it has two advantages
for the study of five-quark components~\cite{An:2013zoa}. First, there is no Pauli
blocking effect for $q$, which would result in lower excitation
energy for the new excitation mechanism of $\Omega^*$ states by
pulling out a $q\bar q$ pair and make the five-quark components
larger in $\Omega^*$ than in $N^*$, $\Lambda^*$ etc. Secondly it
simplifies the model calculation of the five quark system. Only after we could understand the $\Omega^*$ or similarly $\Omega_c^*$ spectroscopy, we may really understand more complicated $N^*$, $\Lambda^*$ etc.

In summary, the LHCb's discovery of the $P_c$ and $P_{cs}$ states is extremely important. For the first time after half century's hard efforts we have finally got experimental observed pentaquark states to check our theoretical models for the pentaquark spectroscopy. It provides a breakthrough point for understanding the pentaquark structure, which is crucial for unquenched quark models to describe the whole baryon spectroscopy.    

\bigskip
\noindent {\bf Acknowledgments:}
This work is supported by the NSFC and the Deutsche Forschungsgemeinschaft (DFG, German Research
Foundation) through the funds provided to the Sino-German Collaborative
Research Center TRR110 “Symmetries and the Emergence of Structure in QCD”
(NSFC Grant No. 12070131001, DFG Project-ID 196253076 - TRR 110), by the NSFC 
Grant No.11835015, No.12047503, and by the Chinese Academy of Sciences (CAS) under Grant No.XDB34030000.



\begin{thebibliography}{}
%Journal Author, Journal \textbf{Volume}, page numbers (year)
%Book Author, \textit{Book title} (Publisher, place, year) page numbers
% etc
%

%\cite{Aaij:2015tga}
\bibitem{Aaij:2015tga} 
  R.~Aaij {\it et al.} [LHCb Collaboration],
%  %``Observation of $J/\psi p$ Resonances Consistent with Pentaquark States in $\Lambda_b^0 \to J/\psi K^- p$ Decays,''
  Phys.\ Rev.\ Lett.\  {\bf 115}, 072001 (2015)
  doi:10.1103/PhysRevLett.115.072001
  [arXiv:1507.03414 [hep-ex]].

%\cite{Aaij:2019vzc}
\bibitem{Aaij:2019vzc} 
  R.~Aaij {\it et al.} [LHCb Collaboration],
  %``Observation of a narrow pentaquark state, $P_c(4312)^+$, and of two-peak structure of the $P_c(4450)^+$,''
  Phys.\ Rev.\ Lett.\  {\bf 122}, no. 22, 222001 (2019)
  doi:10.1103/PhysRevLett.122.222001
  [arXiv:1904.03947 [hep-ex]].

%\cite{Aaij:2020gdg}
\bibitem{Aaij:2020gdg}
%R.~Aaij \textit{et al.} [LHCb Collaboration],
L. Zhang, LHCb collaboration,
%``Evidence of a $J/\psi \varLambda$ structure and observation of excited $\varXi^-$ states in the $\varXi_b^-\to J/\psi \varLambda K^-$ decay,''
Science Bulletin (2021), doi: https://doi.org/10.1016/j.scib.2021.02.030
[arXiv:2012.10380 [hep-ex]].

%\cite{Wu:2010jy}
\bibitem{Wu:2010jy}
  J.~J.~Wu, R.~Molina, E.~Oset and B.~S.~Zou,
  %``Prediction of narrow $N^*$ and $\Lambda^*$ resonances with hidden charm above 4 GeV,''
  Phys.\ Rev.\ Lett.\  {\bf 105}, 232001 (2010)
%  doi:10.1103/PhysRevLett.105.232001
  [arXiv:1007.0573 [nucl-th]].

%\cite{Wang:2011rga} 4g
\bibitem{Wang:2011rga}
  W.~L.~Wang, F.~Huang, Z.~Y.~Zhang and B.~S.~Zou,
  %``$\Sigma_c \bar{D}$ and $\Lambda_c \bar{D}$ states in a chiral quark model,''
  Phys.\ Rev.\ C {\bf 84}, 015203 (2011).
%  doi:10.1103/PhysRevC.84.015203
  [arXiv:1101.0453 [nucl-th]].

%\cite{Yang:2011wz}
\bibitem{Yang:2011wz}
  Z.~C.~Yang, Z.~F.~Sun, J.~He, X.~Liu and S.~L.~Zhu,
  %``The possible hidden-charm molecular baryons composed of anti-charmed meson and charmed baryon,''
  Chin.\ Phys.\ C {\bf 36}, 6 (2012)
%  doi:10.1088/1674-1137/36/1/002, 10.1088/1674-1137/36/3/006
  [arXiv:1105.2901 [hep-ph]].

%\cite{Wu:2012md}
\bibitem{Wu:2012md}
  J.~J.~Wu, T.-S.~H.~Lee and B.~S.~Zou,
  %``Nucleon Resonances with Hidden Charm in Coupled-Channel Models,''
  Phys.\ Rev.\ C {\bf 85}, 044002 (2012)
%  doi:10.1103/PhysRevC.85.044002
  [arXiv:1202.1036 [nucl-th]].

%\cite{Xiao:2021rgp}
\bibitem{Xiao:2021rgp}
C.~W.~Xiao, J.~J.~Wu and B.~S.~Zou,
%``Molecular nature of $P_{cs} (4459)$ and its heavy quark spin partners,''
Phys. Rev. D \textbf{103}, no.5, 054016 (2021)
doi:10.1103/PhysRevD.103.054016
[arXiv:2102.02607 [hep-ph]].

%\cite{Guo:2017jvc}
\bibitem{Guo:2017jvc}
F.~K.~Guo, C.~Hanhart, U.~G.~Mei\ss{}ner, Q.~Wang, Q.~Zhao and B.~S.~Zou,
%``Hadronic molecules,''
Rev. Mod. Phys. \textbf{90}, no.1, 015004 (2018)
%doi:10.1103/RevModPhys.90.015004
[arXiv:1705.00141 [hep-ph]].

%\cite{Chen:2016qju}
\bibitem{Chen:2016qju}
H.~X.~Chen, W.~Chen, X.~Liu and S.~L.~Zhu,
%``The hidden-charm pentaquark and tetraquark states,''
Phys. Rept. \textbf{639}, 1-121 (2016)
%doi:10.1016/j.physrep.2016.05.004
[arXiv:1601.02092 [hep-ph]].

%\cite{Dalitz:1959dn}
\bibitem{Dalitz:1959dn} 
  R.~H.~Dalitz and S.~F.~Tuan,
  %``A possible resonant state in pion-hyperon scattering,''
  Phys.\ Rev.\ Lett.\  {\bf 2}, 425 (1959).
  doi:10.1103/PhysRevLett.2.425
  
%\cite{Kaiser:1995eg}
\bibitem{Kaiser:1995eg}
  N.~Kaiser, P.~B.~Siegel and W.~Weise,
  %``Chiral dynamics and the low-energy kaon - nucleon interaction,''
  Nucl.\ Phys.\ A {\bf 594} (1995) 325
  doi:10.1016/0375-9474(95)00362-5
  [nucl-th/9505043].
  %%CITATION = doi:10.1016/0375-9474(95)00362-5;%%
  
%\cite{Liu:2005pm}
\bibitem{Liu:2005pm}
  B.~C.~Liu and B.~S.~Zou,
  %``Mass and K Lambda coupling of N*(1535),''
  Phys.\ Rev.\ Lett.\  {\bf 96} (2006) no.4,  042002
  doi:10.1103/PhysRevLett.96.042002
  [nucl-th/0503069].
  %%CITATION = doi:10.1103/PhysRevLett.96.042002;%%
  
%\cite{Dong:2021juy}
\bibitem{Dong:2021juy}
X.~K.~Dong, F.~K.~Guo and B.~S.~Zou,
%``A survey of heavy-antiheavy hadronic molecules,''
Progr. Phys. \textbf{41}, 65-93 (2021)
doi:10.13725/j.cnki.pip.2021.02.001
[arXiv:2101.01021 [hep-ph]].

%\cite{Dong:2020hxe}
\bibitem{Dong:2020hxe}
X.~K.~Dong, F.~K.~Guo and B.~S.~Zou, 
%``Why there are many threshold structures in hadron spectrum with heavy quarks,''
Phys.\ Rev.\ Lett.\  {\bf 126} (2021) in press. 
[arXiv:2011.14517 [hep-ph]].

%\cite{Brodsky:1996hc}
\bibitem{Brodsky:1996hc}
S.~J.~Brodsky and B.~Q.~Ma,
%``The Quark / anti-quark asymmetry of the nucleon sea,''
Phys. Lett. B \textbf{381}, 317-324 (1996)
doi:10.1016/0370-2693(96)00597-7
[arXiv:hep-ph/9604393 [hep-ph]].
 
%\cite{Zou:2005xy}
\bibitem{Zou:2005xy}
B.~S.~Zou and D.~O.~Riska,
%``The s anti-s component of the proton and the strangeness magnetic moment,''
Phys. Rev. Lett. \textbf{95}, 072001 (2005)
doi:10.1103/PhysRevLett.95.072001
[arXiv:hep-ph/0502225 [hep-ph]].

%\cite{An:2013zoa}
\bibitem{An:2013zoa}
C.~S.~An, B.~C.~Metsch and B.~S.~Zou,
%``Mixing of the low-lying three- and five-quark $\Omega$ states with negative parity,''
Phys. Rev. C \textbf{87}, no.6, 065207 (2013)
doi:10.1103/PhysRevC.87.065207
[arXiv:1304.6046 [hep-ph]].

 
\end{thebibliography}
\end{document}